\documentclass[pra,10pt,onecolumn,showpacs]{revtex4}%
\usepackage{amsfonts}
\usepackage{amsmath}
\usepackage{amssymb}
\usepackage{graphicx}
\usepackage{acronym}%
\begin{document}
\title{Solution to Satisfiability problem by a complete Grover search with trapped ions}
\author{W. L. Yang$^{1,2},$ H. Wei$^{1,2},$ F. Zhou$^{1,2},$ W. L. Chang$^{3}$}
\email{changwl@cc.kuas.edu.tw}
\author{M. Feng$^{1}$}
\email{mangfeng@wipm.ac.cn}
\affiliation{$^{1}$State Key Laboratory of Magnetic Resonance and Atomic and Molecular
Physics, Wuhan Institute of Physics and Mathematics, Chinese Academy of
Sciences, Wuhan 430071, China }
\affiliation{$^{2}$Graduate School of the Chinese Academy of Sciences, Beijing 100049, China}
\affiliation{$^{3}$Department of Computer Science and Information Engineering, National
Kaohsiung University of Applied Sciences, Kaohsiung 80778, China}

\pacs{03.67.Lx, 89.20.Ff, 03.67.Ac}

\begin{abstract}
The main idea in the original Grover search (Phys. Rev. Lett. \textbf{79}, 325
(1997)) is to single out a target state containing the solution to a search
problem by amplifying the amplitude of the state, following the Oracle's job,
i.e., a black box giving us information about the target state. We design
quantum circuits to accomplish a complete Grover search involving both the
Oracle's job and the amplification of the target state, which are employed to
solve Satisfiability (SAT) problems. We explore how to carry out the quantum
circuits by currently available ion-trap quantum computing technology.

\end{abstract}
\maketitle

\section{INTRODUCTION}

Quantum algorithms, such as the well-known Shor's factoring algorithm
\cite{Shor} and Grover's search algorithm \cite{Grover}, have shown
exponential and quadratic speed-up, respectively, in computation over
classical counterpart, due to the capabilities exploiting the parallelism of
quantum mechanics or interference effects. It is considered that the future
quantum computers should be able to solve some classically intractable
problems \cite{Shor, Hogg1}, e.g., the nondeterministic polynomial-complete
(NPC) problems \cite{Cook1, Garey}, among which the random Boolean
$K$-Satisfiability ($K$-SAT) problem \cite{Cook2} is a central issue in
computer science. Many proposals have so far focused on the solution of hard
instances of $K$-SAT problems by the method of so-called DPLL algorithm
\cite{Davis}, quantum adiabatic algorithm \cite{Hogg2, Ralf, Marko, Kny},
Grover's search algorithm \cite{Ju}, Hogg's algorithm \cite{Peng}, local
search algorithms \cite{Rame, Bar1, Frey, Seitz}, statistical mechanics
approach \cite{Bar2}, etc.

We will concentrate in the present paper on solution to $K$-SAT problems by
Grover search algorithm. Our motivation is to show the possibility with Grover
search finding answers to some solvable $K$-SAT problems. Although most
$K$-SAT, even if $K=3,$ are NP problems and Grover search does not own
exponential speed-up capability in solution, we will consider some solvable
cases of $K$-SAT as examples. As we know from the original Grover's papers and
other subsequent work, Grover search was only considered as a method to
efficiently single out the answer states (or say, target state in the language
of Grover search), whereas the job to find the target state is resorted to a
black box named Oracle. To work out a realistic problem, however, we should
have to consider how to take the Oracle's work. So it naturally arises a
question: Is it possible to design a scheme with consideration of both the job
by Oracle and the job of amplitude amplification?

We have noticed a recent publication \cite{Ju} with quantum circuits for a
complete Grover search including Oracle's job and amplification of target
state, in which a three-variable (i.e., three-qubit in quantum computing (QC)
treatment) SAT problem is solved by employing seven auxiliary qubits and tens
of single and multiple quantum gates. Enlightening by this idea, we will
design some simplified quantum circuits for solutions to some $K$-SAT
problems, which enables us to work with trapped-ion QC.

The ion-trap system \cite{Mon, Cirac, Plenio} favors QC owing to long
coherence time of qubits, high efficiency of detection, and full
controllability of operations. In the context of trapped-ion QC, there have
been some schemes \cite{Fuji, Feng} for implementation of Grover search by
constructing multiqubit conditional operations based on Cirac-Zoller (CZ) gate
\cite{Zoller} or M\o lmer-S\o rensen gate \cite{Molmer}. Experimentally,
two-qubit Grover search has been carried out by $^{111}Cd^{+}$ \cite{Brick}$,$
and eight ions have been confined stably in entanglement in the trap
\cite{eightion}, which is the most qubits ready for QC among the currently
available QC candidate systems. Most recently, Grover search was carried out
by a relatively simple way \cite{Ivan1} making use of collective states of
trapped ions \cite{Ivan2}.

We will explore below the experimental feasibility of solving $K$-SAT problems
by a complete Grover search with trapped ions, where the Grover search will be
implemented either by sequences of single-qubit and multiqubit quantum gates
based on the light-shift (LS) gates \cite{Jonathan} or directly by multiqubit
conditional phase flip (CPF) gates \cite{Yang}. We will compare the two
methods for implementation of our solutions. The outline of the paper is as
follows. In Section II, the $K$-SAT problem and Grover search algorithm are
briefly introduced. In Section III, we apply the complete Grover search to
some satisfiable $K$-SAT problems by quantum circuits, which compared to in
\cite{Ju}, is much simplified with reduced number of the auxiliary qubits and
gating. Then we will discuss the experimental feasibility with trapped ions
for implementation in Section IV. The last section is for our conclusion.

\section{$K$-SAT PROBLEM AND GROVER SEARCH ALGORITHM \textit{ }}

Let us first of all review the $K$-SAT problem briefly. As a paradigmatic
example of a NPC problem, the well-known $K$-SAT problem is actually a
combinatorial search problem in theoretical computer science \cite{explain}.
Generally speaking, the $K$-SAT problem could be expressed by a logical
configuration involving $\xi$ Boolean variables $b_{i}$ ($i=1,2,\cdot
\cdot\cdot,\xi$) with the values $0=FALSE$ and $1=TRUE$. A $K$-SAT formula
$F_{K}$, which consists of $m$ clauses $\{C_{\mu}\}_{\mu=1,2,\cdot\cdot\cdot
m},$ could be written as \cite{Marko},%
\begin{equation}
F_{K}=C_{1}\wedge C_{2}\wedge C_{3}\cdot\cdot\cdot\wedge C_{m},
\end{equation}
where $\wedge$ means logical AND gate, and each clause $C_{i}$ contains a
number of logical variables $b_{i}$ or its negation $\bar{b}_{i}$ (i.e.,
logical NOT gate on $b_{i}$). These variables ($b_{i}$ or $\bar{b}_{i}$) are
connected with each other by $\vee$ (logical OR gate), and the maximal number
of variables in each clause is $K.$ For instance, $F_{2}=(\bar{b}_{2}\vee
b_{4})\wedge(\bar{b}_{2}\vee b_{3})\wedge(b_{1}\vee b_{3})$ is a $2$-SAT
formula with 3 clauses and 4 logical variables. A solution to $F_{K}$ is to
find the values of the logical variables satisfying all clauses simultaneously
in order to make $F_{K}$ be $TRUE$. In some cases, there is no solution to a
SAT problem, which is called unsatisfiable, and in some other cases, there are
possibilities to have multiple solutions to a SAT problem. In the present
scheme, for simplicity, we will restrict our study, exclusively, on the
formulae with only one solution.

The primary idea of Grover's algorithm in \cite{Grover} is to boost the
probability amplitude of the target state so that we could measure the target
state with a considerably high probability. Generally speaking, provided that
the initial state of the system has been prepared in an average superposition
state $\left\vert \Psi_{0}\right\rangle =(1/\sqrt{N})\sum_{i=0}^{N-1}%
\left\vert i\right\rangle $ ($\allowbreak N=2^{n}$ with $n$ the qubit number),
Grover search can be depicted as the iterative operation $G=$ $\hat{D}%
^{(n)}I_{\tau}$ (defined later) by at least $\pi\sqrt{N}/4$ times for finding
the marked state $\left\vert \tau\right\rangle $ with an optimal probability
\cite{Yama,Yang2}, where the quantum phase gate $I_{\tau}=I-2\left\vert
\tau\right\rangle \left\langle \tau\right\vert $ (with $I$ being the identity
matrix) plays an important role of \textit{Selective-Inversion }(SI) to invert
the amplitude of the target state, and the diffusion transform $\hat{D}^{(n)}$
defined as $\hat{D}_{ij}^{(n)}=2/N-\delta_{ij}\ $\ ($i,j=1,2,3,$%
\textperiodcentered\textperiodcentered\textperiodcentered\textperiodcentered
\textperiodcentered\textperiodcentered$\allowbreak N$) is called
\textit{Inversion-About-Average }(IAA).

\section{SOLUTION TO $K$-SAT PROBLEMS BY COMPLETE GROVER SEARCH}

In this section, we will show how Grover search can be carried out to find a
satisfiable solution to the $K$-SAT problem. For simplicity, we first consider
a $2$-SAT formula $F_{2}$ involving three clauses and two Boolean variables,%
\begin{equation}
F_{2}=(\bar{a}\vee\bar{b})\wedge(a\vee b)\wedge a.
\end{equation}
To solve\ $F_{2},$ we have to employ some basic quantum logical gates
including Controlled-NOT(CNOT), AND, and OR, as depicted in Fig. 1 where the
multiqubit CNOT gate $C_{NOT}^{n}$ involves $(n-1)$ qubits as control and the
$n$th qubit as target. The $f-C_{NOT}$ (called function-CNOT) gate inverts the
target state on the condition that the control states satisfy a specific
Boolean function. The $f-C_{NOT}$ could be used for
\textit{eigenvalue-kickback} effect to induce $\pi$ phase shifts on some
component states if the auxiliary state is initially prepared as $(\left\vert
0\right\rangle -\left\vert 1\right\rangle )/\sqrt{2}$ (see Appendix).

The logic operation $AND$ is carried out by a single $C_{NOT}^{3}$
\cite{Vedral}, and implementation of $OR$ consists of a single $C_{NOT}^{3}$
and some $NOT$ operations. To accomplish each of them, an auxiliary qubit,
initially prepared in $|0\rangle,$ is employed to store the output of each operation.

Using above-mentioned logical gates, we depict the quantum circuits in Fig. 2
for a complete Grover search to solve the $2$-SAT problem in Eq. (2). We can
find that, besides two qubits $a$ and $b$ encoding the two variables,
respectively, four auxiliary qubits are required, where $q_{1}$ and $q_{2}$
are to store the results of $\bar{a}\vee\bar{b}$ and $a\vee b,$ respectively.
$q_{3}$ will record the result of $F_{2},$ and $q_{4}$ is the ancilla to
do\ \textit{eigenvalue-kickback}. The first part of the quantum circuit
$\tilde{U}_{0}$ is for Oracle's job, including solution to $F_{2}$ and the
SI\textit{ }operation $I_{\tau}=I-2\left\vert \tau\right\rangle \left\langle
\tau\right\vert $ using the effect of \textit{eigenvalue-kickback}. It
consists of following five steps: (i) gates $1$ and $2$ perform the first
clause $(\bar{a}\vee\bar{b})$ in Eq.(2); (ii) gates $3$, $4$, and $5$
implement the second clause $(a\vee b)$; (iii) the gate $6$ is to restore the
states of the qubits $a$ and $b$; (iv) the gate $7$ executes $AND$ for three
clauses to obtain the value of $F_{2}$, which achieves%
\begin{equation}
(1/2\sqrt{2})(|00\rangle_{ab}|100\rangle_{q_{1}q_{2}q_{3}}+|01\rangle
_{ab}|110\rangle_{q_{1}q_{2}q_{3}}-|10\rangle_{ab}|111\rangle_{q_{1}q_{2}%
q_{3}}+|11\rangle_{ab}|010\rangle_{q_{1}q_{2}q_{3}})(|0\rangle-|1\rangle
)_{q_{4}}.
\end{equation}
Therefore, the target state is $|10\rangle_{ab}|111\rangle_{q_{1}q_{2}q_{3}%
}(|0\rangle-|1\rangle)_{q_{4}},$ which implies the SAT solution to be $a=1$
and $b=0.$ To efficiently single out the target state, we have to discard the
auxiliary qubits $q_{1-3}$, but keep the phase flipping in the qubit subspace.
To this end, we employ reverse operations in an additional circuit $\tilde
{U}_{add}$ to do $\tilde{U}_{0}^{-1}$. After the gating $1^{^{\prime}},$ we
obtain $(1/2)(|00\rangle+|01\rangle-|10\rangle+|11\rangle)_{ab}|000\rangle
_{q_{1}q_{2}q_{3}}(|0\rangle-|1\rangle)_{q_{4}}.$ Consequently, all the
auxiliary qubits are decoupled from the qubits. We may discard $q_{1},$
$q_{2}$ and $q_{3},$ but keep $q_{4}$ for later use.

The function of IAA performed in the subsequent dotted-box of the quantum
circuit is to single out the target state. As mentioned previously
\cite{Feng,Yama,Yang2}, in the implementation of IAA, the diffusion transform
$\hat{D}^{(n)}$ is always unchanged no matter which state is to be searched.
As a result, we may design the IAA as a fixed module. Since the target state
is labeled by SI operation, once IAA (i.e., gate 8) is carried out, we should
be able to single out the target state.

Specifically, the operation in the step of IAA\textit{ }could be represented
by%
\begin{equation}
\hat{D}^{^{(2)}}=-H^{\otimes2}\sigma_{x,b}\sigma_{x,a}C_{PF}^{(2)}\sigma
_{x,a}\sigma_{x,b}H^{\otimes2},
\end{equation}
where $\hat{D}^{^{(2)}}$ is the diffusion transform $\hat{D}^{(n)}$ in the
case of two qubits, $C_{PF}^{(2)}=$diag$\{1,1,1,-1\}$ in the subspace spanned
by \{$\left\vert 0\right\rangle _{a}$, $\left\vert 1\right\rangle _{a}$,
$\left\vert 0\right\rangle _{b}$, $\left\vert 1\right\rangle _{b}$\} inverts
the state $\left\vert 11\right\rangle $ to be $-\left\vert 11\right\rangle ,$%
\begin{equation}
H^{\otimes2}=\prod\nolimits_{i=1}^{2}H_{i}=\frac{1}{2}%
\begin{bmatrix}
1 & 1\\
1 & -1
\end{bmatrix}
\otimes%
\begin{bmatrix}
1 & 1\\
1 & -1
\end{bmatrix}
,
\end{equation}
and $\sigma_{x,i}$ is the single-qubit NOT gate acting on the qubit $i$ ($i=a$
or $b$)$.$

With more variables, the accomplishment of our scheme for solution to $K$-SAT
problem\ would be more complicated. For example, we have designed a quantum
circuit in Fig. 3 for a $2$-SAT formula with three clauses and three
variables, i.e.,%
\begin{equation}
F_{2}^{\prime}=(a\vee b)\wedge(\bar{a}\vee c)\wedge\bar{b},
\end{equation}
where $a$, $b$, and $c$ denote different variables. Due to similarity to
operations in Fig. 2, we will not discuss it in detail. In addition, we have
also designed a quantum circuit in Fig. 4 for a $3$-SAT problem with four
clauses and three variables, which reads%
\begin{equation}
F_{3}=(a\vee b\vee c)\wedge(a\vee\bar{b}\vee c)\wedge b\wedge\bar{c}.
\end{equation}
It is evident that more operations are needed with more variables involved,
and for more clauses, more auxiliary qubits are required. One thing we have to
mention is that, Grover search for more than two qubits are carried out with
probability. So to obtain the final result, we have to perform SI and IAA
repeatedly. But for clarity, we have only plotted in Figs. 4 and 5 a single iteration.

\section{EXPERIMENTAL\ FEASIBILITY WITH TRAPPED IONS BY TWO METHODS}

In the previous section, we have designed some quantum circuits for a complete
Grover search to solve $K$-SAT problems, where the two-qubit $C_{NOT}^{2}$
gate and multiqubit $C_{NOT}^{n}$ ($n\geq3$) gate play crucial roles. To have
efficient quantum operations, however, we expect less operations and shorter
implementing time. So in contrast with conventional methods \cite{Ben,Slea}
using a number of single-qubit and two-qubit gates to compose a $C_{NOT}^{n}$
gate, a straightforward performance of $C_{NOT}^{n}$ gate seems more
attractive \cite{Yang,Wang}. In what follows, we will try to carry out those
quantum circuits with ultracold trapped ions by two different methods for
$C_{NOT}^{n}$ gating: conventional way and straightforward way.

Let us first consider the conventional way by taking LS gates \cite{Jonathan}
as an example. In ion trap QC \cite{Zoller}, the qubits are encoded in each
ion's internal ground state $\left\vert \downarrow\right\rangle $ and excited
state $\left\vert \uparrow\right\rangle $, and quantum gates are made via
excitation of the common vibrational modes. We assume that the ions could be
irradiated individually by lasers, and we have a string of $n$ ions confined
in a linear trap, with strong confinement along $\vec{x}$ and $\vec{y}$
directions, but less strong confinement along the $\vec{z}$ axis. According to
\cite{Jonathan}, conditional on the resonance condition $\Omega_{0}=\omega
_{z}/2$ with $\Omega_{0}$ the Rabi frequency and $\omega_{z}$ the $\vec{z}%
$-axis vibrational frequency, the ac Stark shift induced by the laser in
resonance with the carrier transition frequency allows the ionic internal
state to couple ionic motional state in a way exactly analogous to the red
detuning transition. As a result, the two-qubit $C_{NOT}^{2}$ gate could be
achieved by sequences of laser pulses corresponding to following unitary
operators on $j$th ion: $\hat{C}_{j}^{\{\vartheta,q\}}=\exp[-i\frac{\vartheta
}{2}(\left\vert \uparrow\right\rangle _{j}\left\langle \downarrow\right\vert
+\left\vert \downarrow\right\rangle _{j}\left\langle \uparrow\right\vert )],$
or $\hat{R}_{j}^{\{\vartheta^{^{\prime}},q^{^{\prime}}\}}=\exp[-i\frac
{\vartheta^{^{\prime}}}{2}(a\left\vert \uparrow\right\rangle _{j}\left\langle
\downarrow\right\vert -a^{\dagger}\left\vert \downarrow\right\rangle
_{j}\left\langle \uparrow\right\vert )]$, where $\vartheta=2\Omega_{0}t$,
$\vartheta^{^{\prime}}=2\Omega_{0}\eta t$, $a$ and $a^{\dagger}$ are the
annihilation and creation operators of the quantized center-of-mass (COM)
mode, respectively, and $q(q^{^{\prime}})=I,II$ refers to laser polarization
for achieving the transitions $\left\vert \downarrow\right\rangle
\longrightarrow\left\vert \uparrow\right\rangle $ and $\left\vert
\downarrow\right\rangle \longrightarrow\left\vert aux\right\rangle $ (See Fig.
5), respectively, with $\left\vert aux\right\rangle $ the auxiliary state
defined in \cite{Zoller}.

Concretely speaking, a two-qubit $C_{NOT}^{2}$ gate could be constructed by
following laser sequence: $C_{NOT}^{2}=\hat{R}_{2}^{\{\pi,I\}}\hat{C}%
_{1}^{\{\pi/2,II\}}\hat{R}_{1}^{\{2\pi,II\}}\hat{C}_{1}^{\{\pi/2,II\}}\hat
{R}_{2}^{\{\pi,I\}}\hat{C}_{1}^{^{\prime}}$ including four types of basic
operators: (i) $A$ gate: $\pi/2$ pulse for carrier transition, which
implements single-qubit rotations; (ii) $B$ gate: $\pi$ pulse for red detuning
transition, which changes the electronic state of the ion as well as the
vibrational state of the ion string; (iii) $B^{\ast}$ gate: $2\pi$ pulse for
red detuning transition; (iv) $A^{\ast}$ gate: the last local operation
$\hat{C}_{1}^{^{\prime}}$ in the laser sequence, used for removing the
undesired phase acquired in previous steps. These four types of basic
operators enable us to realize quantum gates between two arbitrary ions in an
ion string mediated by the vibrational mode. Based on $C_{NOT}^{2}$, the
multiqubit $C_{NOT}^{n}$ gate can be constructed by the standard decomposition
scheme \cite{Ben}, where the $C_{NOT}^{n}$ gate (for $n=3,4,...,8$) consists
of $(2^{n}-2)$ $C_{NOT}^{2}$ gates and $2^{n}$ one-qubit gates. In other
words, a $C_{NOT}^{n}$ gate is constituted by $(2^{n+1}-4)$ $A$ gates,
$(2^{n}-2)$ $A^{\ast}$ gates, $(2^{n+1}-4)$ $B$ gates, and $(2^{n}-2)$
$B^{\ast}$ gates, in addition to $2^{n}$ one-qubit gates. So realization of a
$C_{NOT}^{n}$ gate by the conventional way requires $(8\times2^{n}-12)$ laser pulses.

Alternatively, for the straightforward way, we take our recent proposal
\cite{Yang} as an example, in which the qubit encoding in the last ion is
different from that in other ions (See Fig. 5). By this way, we could have a
one-step implementation of CPF gate $C_{PF}^{(n)}$ $(n\geq2).$ According to
the scheme in \cite{Yang}, both success rate and fidelity of the $C_{PF}%
^{(n)}$ gate depend on a parameter $m$ which equals to $\Omega_{\max}%
^{n}/\Omega_{\max}^{i}$ with $\Omega_{\max}^{k}$ $(k=1,2,...,n)$ the maximum
Rabi frequency regarding $kth$ ion by a laser. The $n$th ion is the last ion.
It was shown in \cite{Yang} that the smaller the value of $m$, the better the
performance of $C_{PF}^{(n)}$, and the gate $C_{PF}^{(n)}=$ $J_{11\cdot
\cdot\cdot1}^{(n)}=$diag$\{1,\cdot\cdot\cdot,1,-1\}$ is to make a phase flip
on the $n$th ion, which yields the multiqubit $C_{NOT}^{n}$ gating,%
\begin{equation}
C_{NOT}^{n}=H_{n}C_{PF}^{(n)}H_{n},
\end{equation}
with $H_{n}$ the single-qubit Hadamard gate acting on the $n$th ion. Note that
the single-qubit operation takes negligible time in comparison with multiqubit
$C_{PF}^{(n)}$ gate, so the $C_{NOT}^{n}$ gating time is almost equal to the
$C_{PF}^{(n)}$ gating time, and the realization of a $C_{NOT}^{n}$ gate by a
straightforward way requires only $(n+2)$ individually addressing laser pulses.

In what follows, we compare the two above-mentioned methods working with
trapped $^{40}Ca^{+}$ to carry out the quantum circuits in Figs. 2, 3, and 4.
To have a good confinement of the trapped ions, we have fixed the Lamb-Dicke
parameter to be $\eta=0.02$ throughout our calculation, where $\eta=(\vec
{k}\cos\theta)\sqrt{\hbar/2nM\omega_{z}}$ with $\vec{k}$ the wave vector of
the laser, $\theta$ the angle of the laser radiation with respect to $\vec{z}$
axis, $n$ the number of the ions, and $M$ the mass of the ion. In our
estimate, we\ have set the angle $\theta=30^{\circ}$. As a result, we have
different axial trapping frequencies for different numbers of ions. For
example, in quantum circuits I, II, and III, we have $\omega_{z}/2\pi=2.92$
MHz, $2.50$ MHz, and $1.94$ MHz, respectively. Moreover, in both methods, we
consider strong laser radiation, i.e., $\Omega_{0}=\omega_{z}/2$ in the
conventional way, and $\Omega_{\max}^{i}=\omega_{z}/2$ $(i\neq n)$ in the
straightforward way. For the scheme with LB gates \cite{Jonathan}, the
$C_{NOT}^{2}$ gating time is mainly determined by $B$ and $B^{\ast}$ gating
time since other gates works much faster. So we may only consider the values
of $T_{B}$ and $T_{B^{\ast}}$ to achieve $B$ and $B^{\ast}$ gates with
$T_{B}=\pi/2\Omega_{0}\eta=50\pi/\omega_{z}$, and $T_{B^{\ast}}=2T_{B}.$ In
contrast, for our CPF proposal \cite{Yang}, in each circuit, once $\eta$ is
determined, the $C_{PF}^{(n)}$ gating time is irrelevant to the number of the
qubits involved, but determined by the Rabi frequency regarding the last
qubit, i.e., $\Omega_{\max}^{n}.$ Like in \cite{Yang}, we set $\Omega_{\max
}^{n}=\Omega_{\max}^{i}/10,$ which yields the $C_{PF}^{(n)}$ gating time to be
$T=\pi/\eta\Omega_{\max}^{n}=1000\pi/\omega_{z}$. We have listed our results
in Tab. I, where both methods take millisecond time-scale for an
implementation of our scheme.

In addition, as $C_{PF}^{(n)}$ in the straightforward way has intrinsic
success probability regarding $m=\Omega_{\max}^{n}/\Omega_{\max}^{i}$, we have
also shown by numerical simulation the implementation of the Grover search for
solution to $K$-SAT problem in Fig. 6, in which both the success probability
of the $C_{PF}^{(n)}$ gate and the intrinsic probability of the Grover search
itself are involved. The lower success rates for the three-qubit cases, i.e.,
the solid and dotted-dashed curves in Fig. 6, are due to the intrinsic
probability of the Grover search. Actually, once we have finished the second
iteration, the success rates would be much higher \cite{Yang2}.

\section{DISCUSSION}

To discuss the experimental feasibility of our scheme, we consider ultracold
trapped Calcium ions $^{40}Ca^{+}$ in a linear trap as an example. If we adopt
the conventional way \cite{Jonathan}, the ground state $\left\vert
\downarrow\right\rangle $ and excited state $\left\vert \uparrow\right\rangle
$ could be encoded in $S_{1/2}$ $(m_{j}=-1/2)$ and $D_{5/2}$ $(m_{j}=-1/2)$,
respectively, and $D_{5/2}$ $(m_{j}=-3/2)$ can be one of the candidates for
the auxiliary state $\left\vert aux\right\rangle $. In contrast, once the
straightforward way is utilized, as done in \cite{Yang}, for the first $(n-1)$
ions, the qubits $\left\vert \downarrow\right\rangle $ and $\left\vert
\uparrow\right\rangle $ are encoded in $S_{1/2}$ $(m_{j}=1/2)$ and $S_{1/2}$
$(m_{j}=-1/2)$, namely, the Zeeman sublevels of the ground state $S_{1/2}$
\cite{Oxford}, but for the last $n$th ion, the qubits $\left\vert
\downarrow\right\rangle $ and $\left\vert \uparrow\right\rangle $ are encoded
into $S_{1/2}$ $(m_{j}=1/2)$ and $D_{5/2}$ $(m_{j}=-1/2)$,\ respectively. The
prerequisite of the experiment includes the accurate tuning of the laser
pulses to the desired frequencies and phases, and the initial preparation of
the vibrational mode of the ions to the ground state.

Both the conventional way and the straightforward way include following common
features: i) the COM motion is utilized as the data bus, namely, the quantum
logic operations (except for the carrier transition) involving the degrees of
freedom of the quantized motion. So the vibrational mode should be laser
cooled to the ground state, and thereby heating becomes a dominant source of
decoherence; ii) Individually addressing by lasers is required. As a result,
the laser intensity fluctuation $\Delta\Omega_{0}$ and the phase fluctuation
$\Delta\phi$ should be well controlled for achieving high fidelity of gating
\cite{Fuji, Sch}.

As we know, heating time of the ground vibrational state of the ions in the
linear trap is on timescale of millisec, e.g., longer than $4$ millisec for
multi-ion trapping in NIST experiments \cite{Barrett} and even as long as 190
millisec for a single trapped ion in Innsbruck experiments \cite{Monrev}. In
contrast, the lifetime of the metastable level $D_{5/2}$ of $^{40}Ca^{+}$ is
much long, i.e., for about $1.16$ sec \cite{Lifetime}. So our operational time
should be restricted within tens of millisec. Fortunately, from our
calculation, the required total time $T$ regarding both methods are shorter
than 10 millisec. Nevertheless, to have a completely heating-free operation,
we may accelerate our manipulation by some other ways. One of the ways is to
optimize the quantum circuits \cite{Optimal1}, i.e., minimizing the number of
gates required by using geometric approach \cite{Optimal2}. Alternatively, we
may consider to improve the efficiency of quantum gates. For example, we may
employ a tighter trap to enhance $\omega_{z}$, or consider to exactly adjust
the magic numbers regarding the Lamb-Dicke parameter \cite{Mon2}.

On the other hand, we can also find advantages of the straightforward way over
the conventional way. First, the less number of laser pulses makes operation
easier in practice. In the conventional way \cite{Jonathan}, the number of
laser pulses required to accomplish $C_{NOT}^{n}$ gate is much larger than the
counterpart required in the straightforward way. So overhead could be much
reduced in the straightforward way to generate a multiqubit $C_{NOT}^{n}$.
Secondly, the increase of the qubits in straightforward way \cite{Yang} could
improve the fidelity and the success probability of the $C_{PF}^{(n)}$ gate,
which are favorable for a scalable Grover search. Thirdly, the reduction of
the manipulation steps could diminish computational errors.

As shown in Tab. I, however, the straightforward way, although with much fewer
operational steps, takes comparable time to the conventional way. The reason
is that the $C_{PF}^{(n)}$ gating rate in straightforward way mainly depends
on the value of the Rabi frequency $\Omega_{\max}^{n}$ regarding the $n$th ion
by a laser, namely, to meet the condition $\operatorname{erf}[\eta\Omega
_{\max}^{n}t_{0}/\sqrt{\pi}]\rightarrow1$ \cite{Yang}. On the other hand, to
obtain a $C_{PF}^{(n)}$ gating with big enough fidelity and success
probability, we require $m=\Omega_{\max}^{n}/\Omega_{\max}^{i}\ll1,$ which
greatly restricts the value of $\Omega_{\max}^{n}$ and thereby limits the
speed of $C_{PF}^{(n)}$ gate. As a result, in the few-qubit case, the
straightforward way works only slightly faster than the conventional way,
whereas it would be more and more efficient than the conventional way with
more qubits involved. Furthermore, as operational overhead could be much
reduced in the straightforward way, we prefer to use it for implementation of
our scheme even in the few-qubit case.

Another point we should mention is that, the scheme in \cite{Yang} is very
sensitive to $m$. As shown in Fig. 6, with the value of $m$ larger than 0.04,
the laser intensity fluctuation and phase fluctuation would lead to
considerable affect on implementation. To avoid this detrimental influence, we
have to keep the laser in high stability, or we take the case with $m<0.04,$
which, however, would yield a longer gating time. So a trade-off would be
taken in realistic implementation with the straightforward way.

For solving $K$-SAT problems with multiple solutions, we will need more gates
to exclude the target states that have been found previously. As the design of
the circuit for this job is straightforward and strongly relevant to the
specific solution \cite{Ju}, we will not go further along this line in the
present work. But more gates will yields longer time and higher requirement
for implementation, which brings about more challenges.

The solution of $K$-SAT problems with more clauses needs more qubits. With
more than two qubits involved, however, Grover search works only
probabilistically. So several iterations are needed to accomplish a solution.
On the other hand, with more ions confined, the vibrational mode spectrum
becomes more and more complicated and the ions' spacing would be decreasing,
which yields individually addressing of the ions to be more intricate. As a
result, the extension of QC from a few qubits to a large number of qubits is
quite technically challenging. Nevertheless, as eight ultracold trapped ions
in entanglement have been experimentally available \cite{eightion}, our
treatment in quantum circuits I, II, and III, which involves six, seven, and
nine ions, respectively, looks very promising under currently achieved
technology\emph{ }in trapped ion system.

\section{CONCLUSION AND ACKNOWLEDGMENT}

It is still hard to find a practical quantum algorithm. That is why we have
had few working quantum algorithms so far. As one of the most frequently
mentioned quantum algorithms, however, Grover search had never been applied to
any really practical problems, although it was mentioned to be useful for
searching some personal information from a phone book. Most relevant proposals
and experiments so far have neglected the Oracle's work. Actually when
Oracle's job is involved in the implementation, much overhead would be needed,
which makes the implementation of Grover search more complicated, but more
realistic and practical.

The contribution of the present work lies in two aspects. First, simplified
quantum circuits for solutions to $K$-SAT problems are designed. Compared to
\cite{Ju}, we have less auxiliary qubits and less operations in our design to
enable solutions to $K$-SAT problems by less than ten qubits, which is
important in view of experimental realization. Secondly, we have explored the
possibility to accomplish the operations required in the quantum circuits with
trapped ions. Our discussion would be also helpful to apply our scheme to
other QC candidate systems.

In summary, we have concentrated on the application of Grover search algorithm
on some solvable $K$-SAT problems with few ultracold trapped ions.
Specifically, we have designed quantum circuits to solve some 2-SAT and 3-SAT
problems, and presented the feasibility, challenge, and possible efforts to
accomplish the schemes with currently or near future available ion trap
technology\emph{. }We believe that our design could be further optimized and
would be useful for exploring application of QC.

The work is supported by NNSF of China under Grants No. 10774163 and No.
10774042. WLC acknowledges support from NSC under Grants No.
96-2221-E-151-008- and 96-2218-E-151-004-.

\section{APPENDIX: EIGENVALUE KICKBACK EFFECT}

Provided that we prepare the state of the target qubit $q_{c}$ as $(\left\vert
0\right\rangle -\left\vert 1\right\rangle )/\sqrt{2}$ and the quantum state of
the control qubits as $\left\vert \Psi\right\rangle _{0}=\sum\nolimits_{\rho
\in\lbrack A^{^{\prime}}]}c_{\rho}\left\vert \rho\right\rangle +\sum
\nolimits_{\upsilon\in\lbrack B^{^{\prime}}]}c_{\upsilon}\left\vert
\upsilon\right\rangle ,$ where $[A^{^{\prime}}]=\{i\mid f(i)=1,$ $i\in
S=(0,1,\cdot\cdot\cdot,2^{n-1})\}$ is the set of satisfied assignments and
$[B^{^{\prime}}]=S-[A^{^{\prime}}]$ is the complementary set, the $f-C_{NOT}$
gate (Fig. 1(c)) will result in \textit{eigenvalue-kickback, }which\textit{
}could effectively flip the phase of some components (i.e., $\left\vert
\rho\right\rangle $) of the state $\left\vert \Psi\right\rangle _{0}$. This
mechanism can be concretely explained as,%
\begin{align*}
&  f-C_{NOT}\{(\sum\nolimits_{\rho\in\lbrack A^{^{\prime}}]}c_{\rho}\left\vert
\rho\right\rangle +\sum\nolimits_{\upsilon\in\lbrack B^{^{\prime}}%
]}c_{\upsilon}\left\vert \upsilon\right\rangle )\otimes(\left\vert
0\right\rangle -\left\vert 1\right\rangle )/\sqrt{2}\}\\
&  =\{\sum\nolimits_{\rho\in\lbrack A^{^{\prime}}]}c_{\rho}\left\vert
\rho\right\rangle \otimes NOT(\left\vert 0\right\rangle -\left\vert
1\right\rangle )+\sum\nolimits_{\upsilon\in\lbrack B^{^{\prime}}]}c_{\upsilon
}\left\vert \upsilon\right\rangle \otimes(\left\vert 0\right\rangle
-\left\vert 1\right\rangle )\}/\sqrt{2}\\
&  =\{\sum\nolimits_{\rho\in\lbrack A^{^{\prime}}]}c_{\rho}\left\vert
\rho\right\rangle \otimes(\left\vert 1\right\rangle -\left\vert 0\right\rangle
)+\sum\nolimits_{\upsilon\in\lbrack B^{^{\prime}}]}c_{\upsilon}\left\vert
\upsilon\right\rangle \otimes(\left\vert 0\right\rangle -\left\vert
1\right\rangle )\}/\sqrt{2}\\
&  =\{-\sum\nolimits_{\rho\in\lbrack A^{^{\prime}}]}c_{\rho}\left\vert
\rho\right\rangle +\sum\nolimits_{\upsilon\in\lbrack B^{^{\prime}}%
]}c_{\upsilon}\left\vert \upsilon\right\rangle \}\otimes(\left\vert
0\right\rangle -\left\vert 1\right\rangle )/\sqrt{2}%
\end{align*}

FIG. 1. The required basic logic gates, where \textbullet\ denotes the control
qubit and $\oplus$ the target qubit. (a) Two-qubit controlled-NOT gate
$C_{NOT}^{2}$, where qubit $1$ and $2$ denote control and target qubits,
respectively; (b) multiqubit controlled-NOT gate $C_{NOT}^{n}$ with $n-1$
control qubits ($q_{_{1}},\cdot\cdot\cdot,q_{n-1}$); (c) the $f-C_{NOT}$
(function-Controlled-NOT) gate, where the Boolean expression $f(q_{1}%
;\cdot\cdot\cdot;q_{n-1})$ can be satisfied by one or more than one Boolean
variable assignment; (d) AND gate; (e) OR gate.

FIG. 2. Quantum circuit for Grover search to solve a $2$-SAT problem in Eq.
(2), where $H$ denotes Hadamard gate, and the three auxiliary qubits $q_{1-3}$
return to their initial states $\left\vert 0\right\rangle $ after the
operations in blocks $\tilde{U}_{0}$ and $\tilde{U}_{add}.$

FIG. 3. Quantum circuit for one iteration in Grover search to solve the
$2$-SAT problem in Eq. (6), where the dots mean omitted operations SI and IAA.

FIG. 4. Quantum circuit for one iteration in Grover search to solve the
$3$-SAT problem in Eq. (7), where the dots mean omitted operations SI and IAA.

FIG. 5. Schematic setup for implementing Grover search in a linear trap, where
$n$ ions are individually addressed by $n$ lasers. The left inset shows the
ionic-level configuration in LB gate \cite{Jonathan}, where the bold lines
mean the levels encoding qubits, and the transition excited by the laser pulse
depends on the laser polarization. The right inset shows the ionic level
configuration in \cite{Yang}, where the qubit encoding in nth ion is different
from in other ions, as labeled by the bold lines.

FIG. 6. The search probability for finding the solution to a $K$-SAT problem
in one iteration in Grover search versus the parameter $m$, where the dotted,
solid, dashed-dotted curves represent the cases of the quantum circuits I, II,
and III, respectively.

\bigskip

\textbf{TABLE I.} Required time for implementation of one iteration in Grover
search on $^{40}Ca^{+}$ by the quantum circuits I (Fig. 2), II (Fig. 3), and
III (Fig. 4) using two different methods based on the LS gates \cite{Jonathan}
(called conventional way) and multiqubit $CPF$ gate \cite{Yang} (called
straightforward way). In the conventional way, the total time $T$ is the
summation of the operational time for $B,$ and $B^{\ast}$ gate, specifically,
$T=\{N[B]+2N[B^{\ast}]\}T_{B},$ with $N[B],$ and $N[B^{\ast}]$ the number of
the gates $B$, and $B^{\ast}$ in the circuits, respectively, and $T_{B}$ the
time for $B$ gating. In the straightforward way, $C_{PF}^{(i)}$ gating time is
irrelevant to the number of the qubits involved, and $N[C_{PF}^{(i)}]$ denotes
the total number of $C_{PF}^{(i)}$ required in the circuits with $i=2,3,4,5$
being the number of the qubits, and we have set $m=0.1$.%

\begin{tabular}
[c]{l|l|llll|lllll}\hline\hline
& $\vec{z}$-axis trap frequency &  & Conventional & Way &  &  &
Straightforward & Way &  & \\
& $\ \ \ \omega_{z}/2\pi$ $(MHz)$ & $\ \ N[B]$ & $\ N[B^{\ast}]$ & $T_{B}$
$(\mu$s) & $T$ (ms) & $N[C_{PF}^{(2)}]$ & $\ \ \ N[C_{PF}^{(3)}]$ &
$N[C_{PF}^{(4)}]$ & $N[C_{PF}^{(5)}]$ & $\ T$ (ms)\\\hline
Circuit I & \ \ \ \ \ \ 2.92 & \ \ \ 118 & \ \ 59 & 8.562 & 2.021 & \ \ \ 1 &
\ \ \ \ \ \ 5 & \ \ \ 2 & \ \ 0 & \ \ 1.370\\
Circuit II & \ \ \ \ \ \ 2.50 & \ \ \ 134 & \ \ 67 & 10.0 & 2.680 & \ \ \ 1 &
\ \ \ \ \ \ 4 & \ \ \ 3 & \ \ 0 & \ \ 1.600\\
Circuit III & \ \ \ \ \ \ 1.94 & \ \ \ 246 & \ \ 123 & 12.887 & 6.340 &
\ \ \ 1 & \ \ \ \ \ \ 8 & \ \ \ 1 & \ \ 2 & \ \ 3.093\\\hline\hline
\end{tabular}

\bigskip\newpage%

\begin{figure}
[ptb]
\begin{center}
\includegraphics[
height=11.361in,
width=7.8923in
]%
{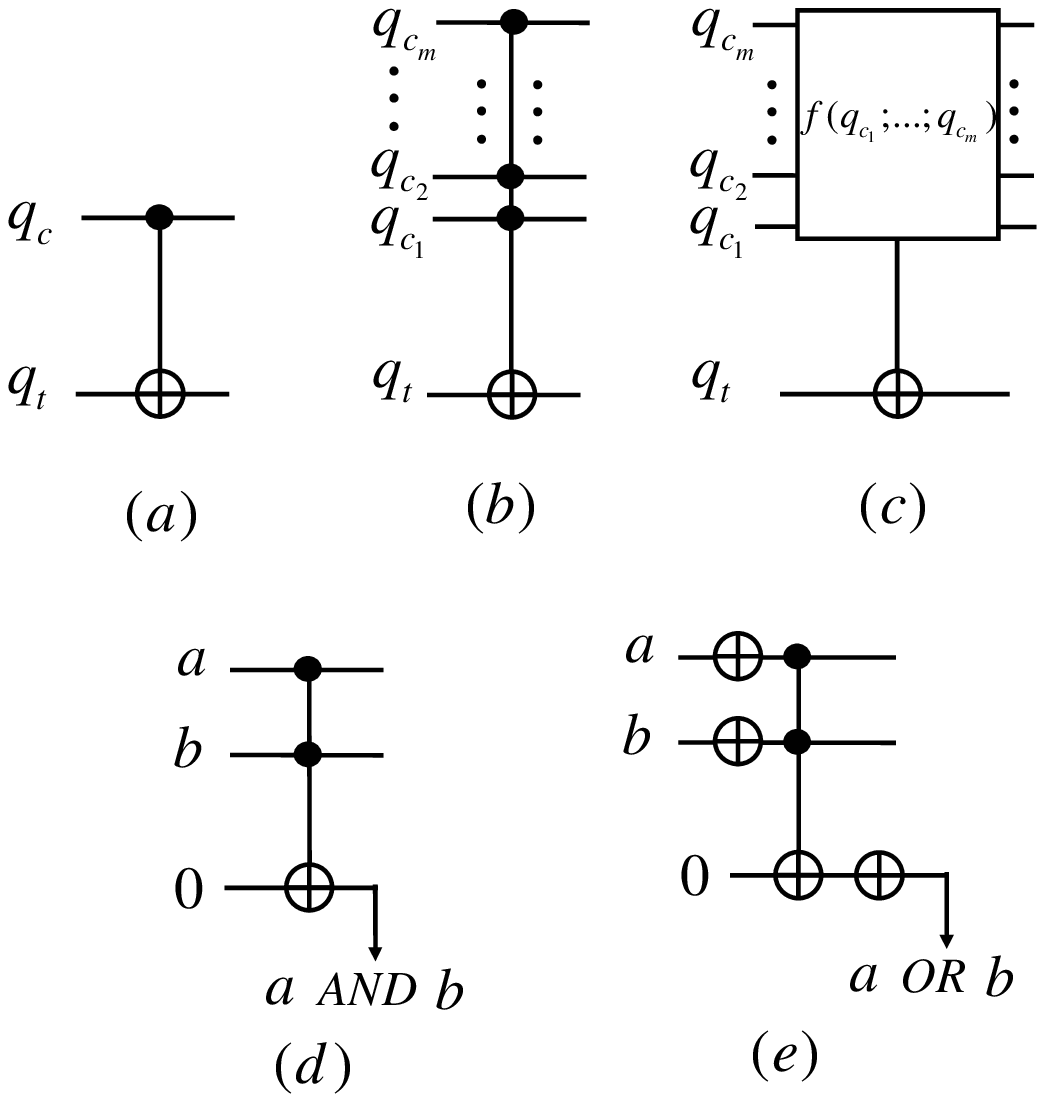}%
\end{center}
\end{figure}
\newpage%

\begin{figure}
[ptb]
\begin{center}
\includegraphics[
height=11.361in,
width=7.8923in
]%
{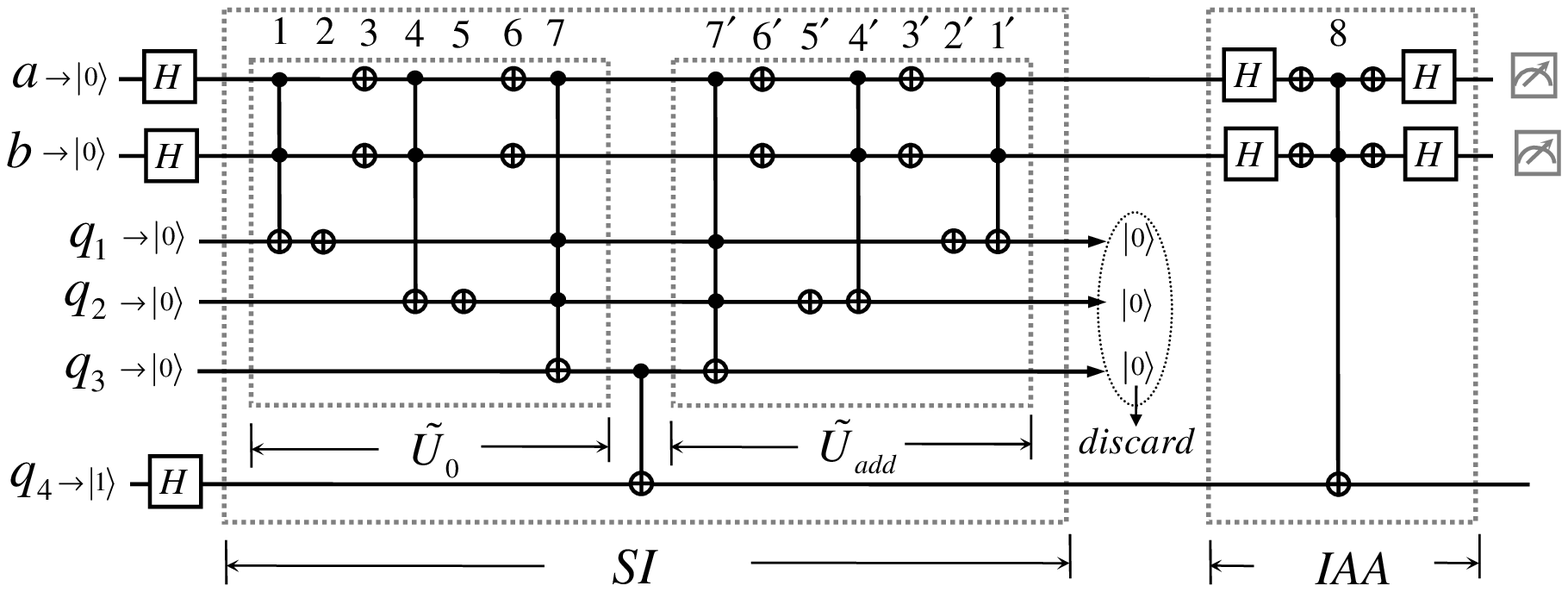}%
\end{center}
\end{figure}

\newpage%

\begin{figure}
[ptb]
\begin{center}
\includegraphics[
height=11.361in,
width=7.8923in
]%
{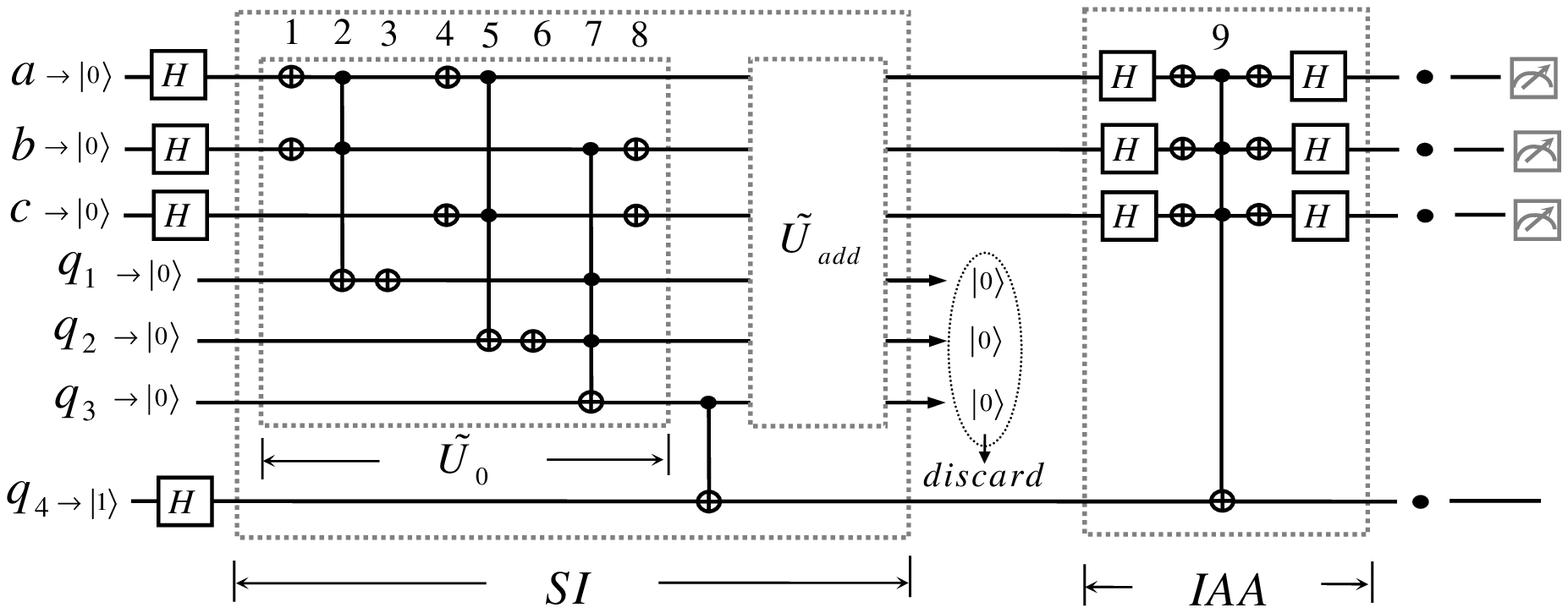}%
\end{center}
\end{figure}
%

\begin{figure}
[ptb]
\begin{center}
\includegraphics[
height=11.361in,
width=7.8923in
]%
{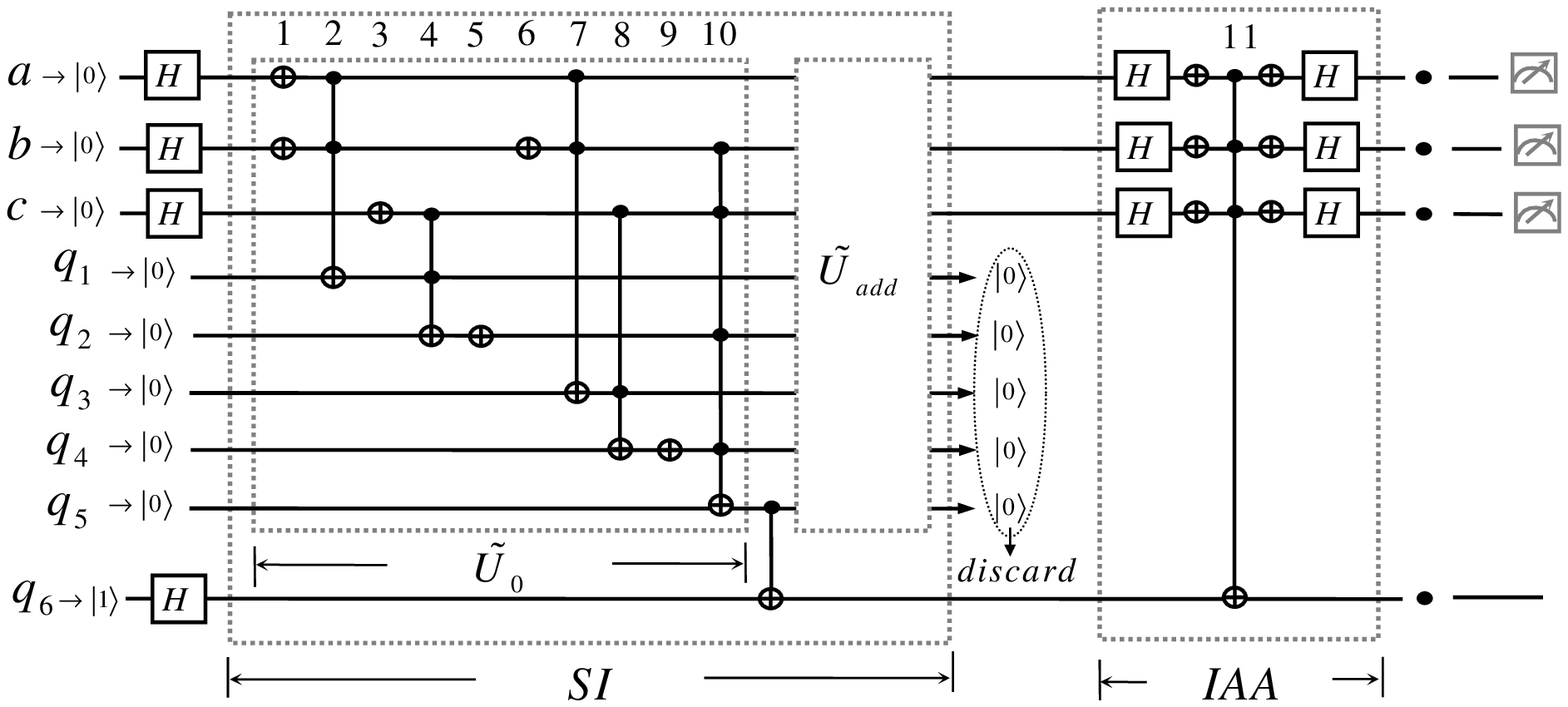}%
\end{center}
\end{figure}

\newpage%

\begin{figure}
[ptb]
\begin{center}
\includegraphics[
height=11.361in,
width=7.8923in
]%
{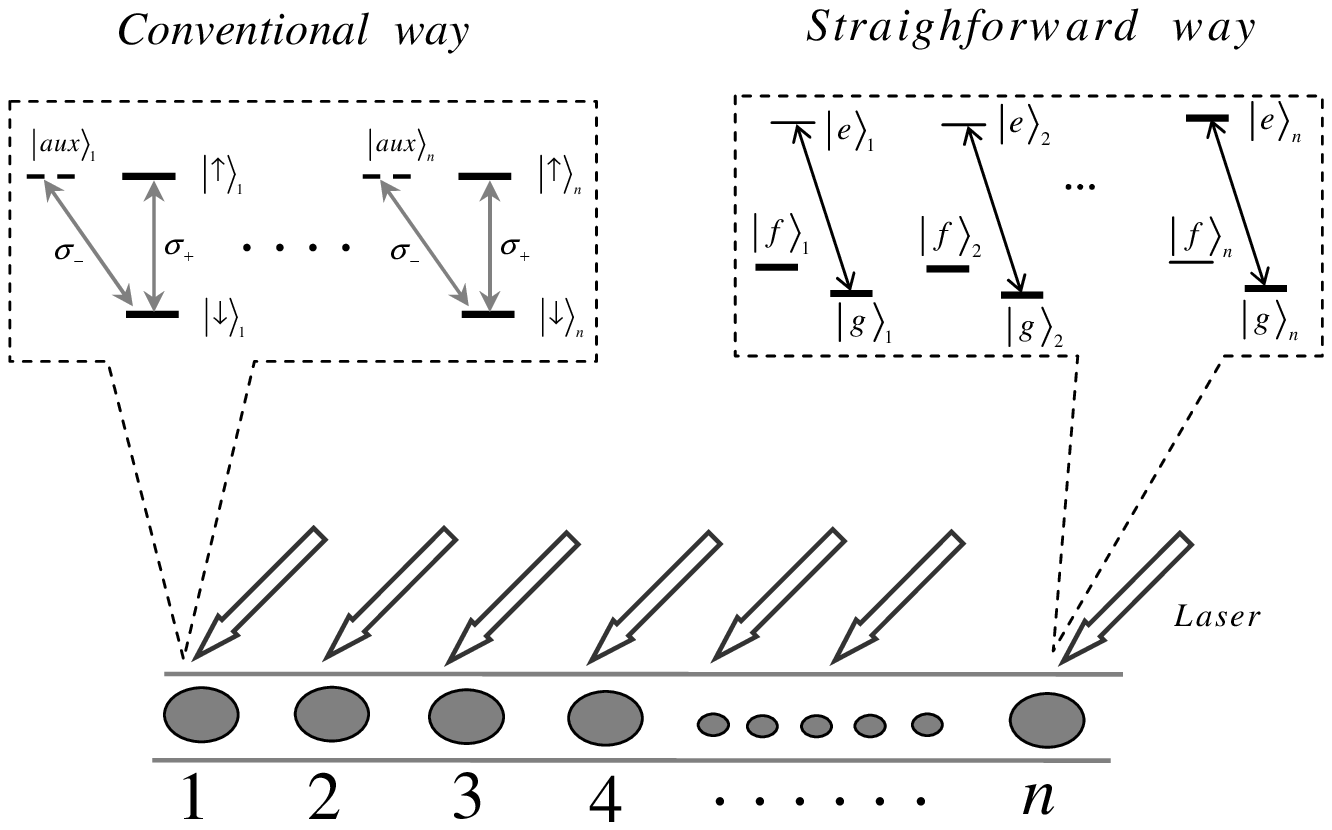}%
\end{center}
\end{figure}

\newpage%

\begin{figure}
[ptb]
\begin{center}
\includegraphics[
height=7.0024in,
width=10.7522in
]%
{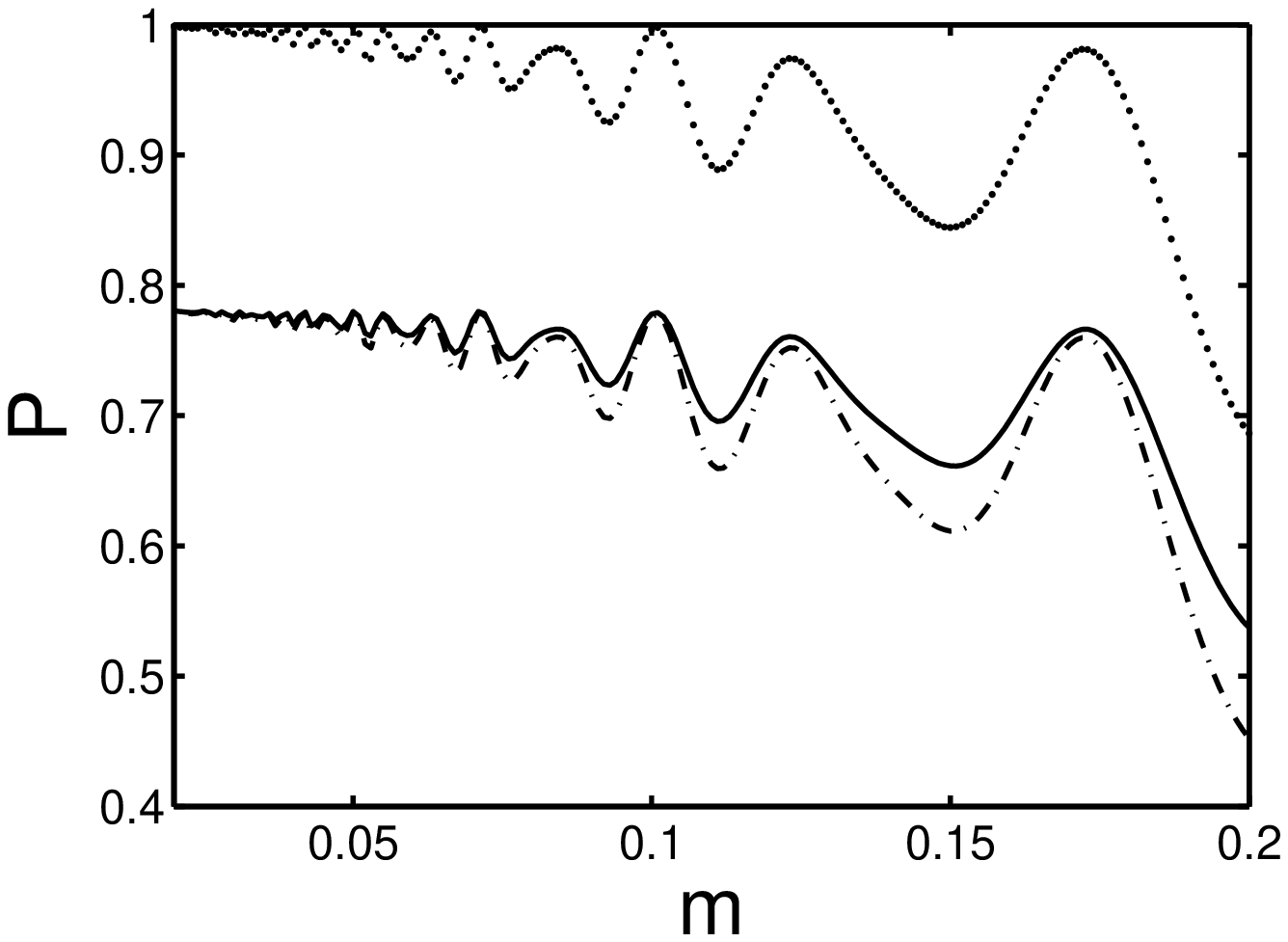}%
\end{center}
\end{figure}


\begin{thebibliography}{99}
\bibitem {explain}Strictly speaking, the $K$-SAT problem becomes NPC class for
$K\geq3$, and 1-SAT and 2-SAT problems with Boolean formula $F_{K}$ involving
at most 2 literals in each clause belong to $P$ class. With increase of the
number of variables, the $K$-SAT problem displays a phase transition from
satisfiable to unsatisfiable phase. It is around this SAT-UNSAT transition
that the typical complexity of the problem raises rapidly, which is beyond the
scope of the present work.
\end{thebibliography}
\end{document}